\documentclass[aps,prl,reprint,floatfix,superscriptaddress]{revtex4-2}

\usepackage{float}
\usepackage{amsmath}
\usepackage{comment}
\usepackage{amsthm}
\usepackage{bbold}
\usepackage{graphicx}
\usepackage{braket}
\usepackage[colorlinks=true, allcolors=blue]{hyperref}

\newcommand\norm[1]{\lVert#1\rVert}
\newcommand{\LL}{\mathcal{L}}

\begin{document}
\title{Krylov space perturbation theory for quantum synchronization in closed systems}
\author{Nicolas Loizeau}
\affiliation{Niels Bohr Institute, University of Copenhagen, Copenhagen, Denmark}
\author{Berislav Bu\v ca}
\affiliation{Niels Bohr Institute, University of Copenhagen, Copenhagen, Denmark}
\affiliation{Universit\'{e} Paris-Saclay, CNRS, LPTMS, 91405, Orsay, France}
\affiliation{Clarendon Laboratory, University of Oxford, Parks Road, Oxford OX1 3PU, United Kingdom}

\date{\today}

\begin{abstract}

Strongly interacting quantum many-body systems are expected to thermalize, however, some evade thermalization due to symmetries. Quantum synchronization provides one such example of ergodicity breaking, but previous studies have focused on open systems. Here, motivated by the problem of ergodicity breaking in closed systems and the study of non-trivial dynamics, we investigate synchronization in a closed disordered Heisenberg spin chain.
In the presence of large random disorder, strongly breaking the permutation symmetry of the system, we observe the emergence of spatial synchronization, where spins lock into locally synchronized patches. This behavior can be interpreted as a fragmentation of the global dynamical symmetry $S^+$ into a collection of local dynamical symmetries, each characterized by a distinct frequency.
In the weak-disorder regime, still without permutation symmetry, we show that the synchronization mechanism can be understood perturbatively within Krylov space. In the absence of disorder, the Krylov space associated with the dynamical symmetry $S^+$ is two-dimensional. Introducing disorder couples this subspace to the remainder of the Krylov space. This coupling leads only to a second-order correction to the frequency of the dynamical symmetry, thereby preserving coherent oscillations despite the presence of small disorder. At stronger disorder, the perturbation modifies $S^+$ so that it acquires a finite lifetime, providing an example of a transient dynamical symmetry.

\end{abstract}

\maketitle

\begin{figure}
\centering
\includegraphics[width=0.4\textwidth]{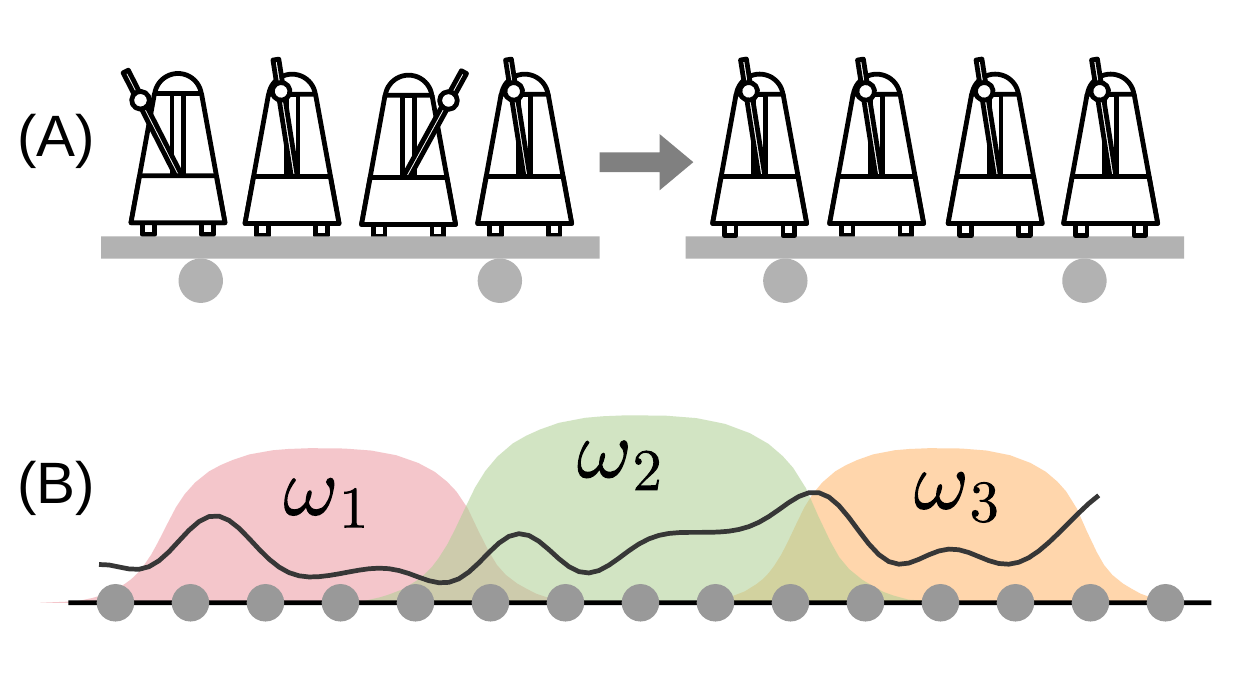}
\caption{(A) A well know example of classical synchronization: when a set of metronomes are coupled together, they end up in phase. (B) Here we are studding a quantum analog. We consider a spin chain with interactions and a disordered magnetic field. We observe that the spins organize in local dynamical symmetries (colors) oscillating at different frequencies.}
\label{fig:drawing}
\end{figure}

{\it Introduction -- } 
Complex dynamics are ubiquitous, and understanding them is essential across the sciences, from biology (e.g. \cite{Gerlich2003, Alon2006}) to economics (e.g. \cite{Eliasson1991, Vinayak2014}). Synchronization is one such striking complex dynamics phenomena, in which interacting systems adjust their motions and behave collectively. It arises in a wide range of many-body contexts, including coupled pendula (Fig \ref{fig:drawing}), neural oscillations in the brain \cite{Fell2011}, epidemic spreading, and power grids \cite{Eroglu03072017}, among others. Explaining synchronization has been a major success of dynamical systems theory and the study of deterministic chaos \cite{BOCCALETTI20021, Pecora1990}.

Intense work in the last decade has extended these classical non-linear results to open quantum systems, mainly e.g., quantum van der Pol oscillators, bosons, or large spin-S systems \cite{Walter2014, Walter2015, Xu2014, Impens2023, Weiss2016, Lee2014, Nigg2018, Mok2020, Eneriz2019, Siwiak2020, Mari2013, Lee2013, Kwasigroch2017, Wachtler2020, Halati2022, Dutta2019, Dubois2021, Bacsi2020, Alaeian2021, Piazza2015, Mivehvar2021, Lin2020, Scarlatella2021, Tao2025, Dutta2025, Vaidya2025, Zhang2025, Schmolke2024, Campbell_2025_dissipation, Aifer2024, Vaidya2024_dissipative, natale2025, Li_2026}. 
These previous works have emphasized open or semiclassical systems with high-dimensional (and continuous) local spaces, and driven and dissipative dynamics. Synchronization in spin chains is an emerging topic \cite{Li2023, Reimann2023, Ghildiyal2025_spin}, but remains largely unexplored for closed systems. This is due to the fact that the notion of synchronization in closed quantum systems has not yet been clearly defined.

Understanding synchronization in spin systems is interesting for experimental applications with spin system platforms. For example, the phenomenon of synchronization blockade \cite{Lorch2017, Tan2022, Kehrer2025_blockade, Nadolny2023, Solanki2023, Tan2022_blockade} has recently been experimentally demonstrated \cite{Koppenhofer2020}. More practically, synchronizing spins strongly in a quantum magnet could allow for highly homogeneous and coherent time-dependent magnetic field sources. Developing such sources has significant potential to improve the resolution of MRI images for which homogeneity and time-dependent coherence of the magnetic field are a major limiting factor \cite{Ladd2018}.

At a more fundamental level, understanding synchronization in closed systems is important because it provides a striking example of ergodicity breaking.
Strongly interacting \emph{closed} systems are expected to thermalize, i.e. relax to stationarity quickly according to the eigenstate thermalization hypothesis (ETH) \cite{Polkovnikov2011, Mori2018, Rigol2008, Buca2023, Doyon2017, Ampelogiannis2023}. Indeed, in a large closed system, thermalization (without invoking any effective or approximate dissipation) is expected to drive the system toward a stationary thermal state rather than a synchronized one, since synchronization is inherently a non-stationary phenomenon. 

Other classes of closed systems are known to avoid thermalization. For example, many-body localized (MBL) systems, due to disorder, avoid it while remaining stationary \cite{Nandkishore2015, lbits1,Abanin2019, Smith2016, Alet2018,DaleyMBL,Gunawardana2022,Iversen_2023}.  
Conversely several systems evade thermalization and instead exhibit persistent non-stationary dynamics, including systems with quantum many-body scars, time crystals, and related phenomena \cite{Buca2020, Else2020, Buca2019, Wu2024, Zaletel2023, Moudgalya_2022, Serbyn2021, Turner2018, Daviet2024, Bull2022, Sarkar2024, Pakrouski2021, Carollo2024, Kardar2025, Jiang2025, Sanada2023, Cabot2024, Cosme2025, Nakanishi2023, schumann2026, Wang2025, Larsen_2024, hosseinabadi2025, Gribben2025, Zhu_2019, Feng2025, Fernandes2025, Kongkhambut2022,Moudgalya2024}. These non trivial dynamics are supported by symmetries, and in particular dynamical symmetries: observables that satisfy $[H, A_\omega]=\omega A_\omega$ \cite{Buca2019,Buca2020attractors, Marko2,Tindall2020, Buca2023,Wachtler2024,Chinzei, Moharramipour2024, Vivek2025_symmetries, Li2026, Buca2022, Srivatsa2020, Majidy2023, Majidy2024}.

An intriguing question is: Can the inherent disorder of MBL be combined with the non-stationary nature of dynamical symmetries in order to induce synchronization in a closed system?

In this Letter we answer this question affirmatively and focus on studying synchronization in closed spin systems through the lens of dynamical symmetries. 

Measures of synchronization which have been previously used for the open system framework were primarily based on phase-locking of correlations, or Husimi Q-functions \cite{HUSIMI, Tindall2020, Roulet2018, zhao2025}. By contrast, we find it convenient and straightforward to rather use natural dynamics of local observables to define synchronization in analogy with classical systems, e.g. see Fig.~\ref{fig:drawing}. 
Crucially, we emphasize again that we go beyond the usual open-system framework, for which such measures were originally developed, both in the Markovian \cite{dutta2025_Markovian, Zhang2024} and non-Markovian \cite{Li2024_non_markovian} cases, and instead focus on the \emph{natural} thermalization inherent to the closed system itself.
We show, based on \cite{Buca2023, Loizeau2025}, that much like closed systems can thermalize, they can also synchronize.

As we will demonstrate by using perturbation theory in Krylov space and dynamical symmetries, the natural thermalization dynamics of a disordered system can lead to all (or most) of the non-synchronized modes dissipating in the long-time limit leaving only one or more synchronized ones.

{\it Setup -- }We are interested in the following disordered Heisenberg spin chain:
\begin{align}
    H =& \sum_i \left( X_i X_{i+1} + Y_i Y_{i+1} + Z_i Z_{i+1} \right)\label{eq:H}\\ 
    &+\sum_i (1+w h_i)X_i\nonumber
\end{align}
where $h_i$ are uniform random numbers in $[-1,1]$ and $w$ is the disorder parameter.
When $w=0$ , the model admits the dynamical symmetry $[H, S^+]=-2S^+$ with $S^+=\sum_j Z_j+iY_j$.

Here we define synchronization as the existence of an extensive translation invariant dynamical symmetry at non zero disorder $w$.
Therefore, we investigate the sensitivity of dynamical symmetry $S^+$ to disorder. Although the disorder explicitly breaks the $S^+$ symmetry, we will show that instead of completely destroying it, it dresses it without significantly changing its frequency for small disorders.

\begin{figure}
\centering
\includegraphics[width=0.48\textwidth]{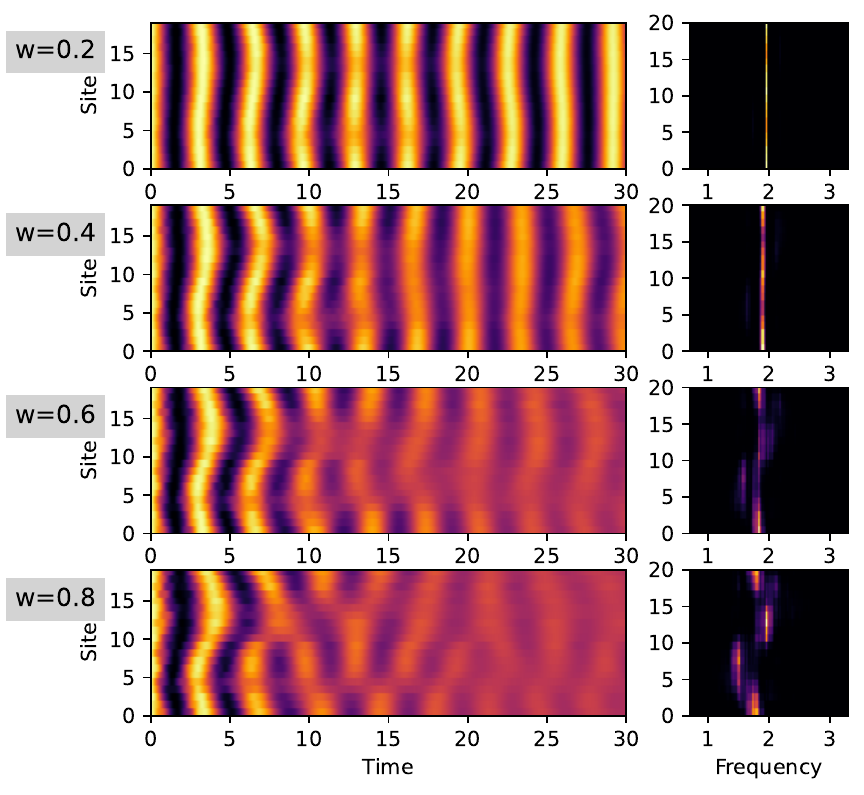}
\caption{Time evolution of $\langle Z_i\rangle$ for a single random realization of Hamiltonian \eqref{eq:H}. The vertical axis is the spacial site index. The right plots show the Fourier transform of the time signal. At low disorder, the spins synchronize. At larger disorder (w=0.8), we observe the formation of local patches of about 5 spins oscillating at different frequencies. }
\label{fig:dynamics}
\end{figure}

First let's look at dynamics under Hamiltonian \ref{eq:H}.
In figure \ref{fig:dynamics} we show the dynamics of $\bra{0}Z_i(t) \ket{0}$ for a single random realization and different values of disorders $w$.
For small disorder, the spins oscillate in phase: they synchronize, for large enough disorder, we observe the formation of different patches oscillating at different frequencies. The spins stay in synchronization with their immediate neighborhood. This is particularly visible in the left panel of figure \ref{fig:dynamics} at $w=0.8$ where, for example, one patch between sites 4 and 10 oscillates at frequency $\sim1.5$ while another patch between sites 10 and 15 oscillates at frequency $\sim2.0$. This phenomenon suggests that the initial dynamical symmetry $S^+$ splits into a sum of dynamical symmetries supported on smaller sectors with each frequency of their own. The extraction of these local dynamical symmetries is discussed in the End Matter (fig \ref{fig:patches}).

{\it Dynamical symmetries in Krylov space -- }
In order to study the existence of dynamical symmetries and their stability with respect to disorder, we will work in Liouvillian Krylov space. These Liouvillian Krylov space techniques have been increasingly used to study integrability and quantum chaos \cite{Loizeau2025, Parker2019, peacock2025, Yates2020, Yates2021, Rabinovici2021, Hörnedal2022, Caputa2022, Erdmenger2023, Camargo2024, Cao2021, Menzler2024, Ballar2022, Balasubramanian2025, Scialchi2025, Yeh2025, Bartsch2024}.  In particular, in ref \cite{Loizeau2025} we have developed a method to probe dynamical symmetries in the thermodynamic limit in Krylov space. The method is closely related to ideas laid out in refs \cite{Teretenkov2025, Uskov2024, shirokov2025, Pinna2025, Fullgraf}. Let's recall it.

The first step of the method is to construct the Krylov space of an observable $O_0$ under Hamiltonian $H$ using Lanczos algorithm.
The algorithm starts with a seed operator $O_0$ then constructs an orthonormal basis of operators by recursively applying the Liouvillian $\mathcal{L}$ to $O_0$ while orthonormalizing at each step \cite{Parker2019, Nandy2024,Claeys}.
The first iteration is given by $O_1 =\LL O_0/b_1 = [H,O_0]/b_1$ and
$b_1=\norm{\LL O_0}$
and for $n>2$,
\begin{align}
    O_{n}' & = \LL O_{n-1}-b_{n-1}O_{n-2}, \nonumber \\
    O_n & = \frac{O_{n}'}{b_{n}}, \nonumber \\
    b_n &= \norm{O_n' }.
    \label{eq:lanczos}
\end{align}
where  $\norm{O}^2 = \frac{1}{2^N}{\rm Tr}[O^2]$.
The algorithm yields an orthonormal `Krylov-basis' $\{O_n\}$ and `Lanczos coefficients' ${b_n}$. 

From the $b_n$ we then construct the Liouvillian in the open Krylov chain

\begin{equation}
\Tilde{\LL}=i\begin{pmatrix}
0   & -b_1   & 0     & \cdots & 0     & 0  \\
b_1   & 0   & -b_2   & \cdots & 0     & 0  \\
0     & b_2   & 0   & \cdots & 0     & 0  \\
\vdots& \vdots& \vdots& \ddots & \vdots\\
0     & 0     & 0     & \cdots & 0  & -b_L \\
0     & 0     & 0     & \cdots & b_L+b_{L+1} & -2b_{L+1}
\end{pmatrix}\label{eq:open_chain}
\end{equation}

where the $b_{L+1}$ terms encode a non hermitian boundary condition that represents the fact that the Krylov chain is infinite by letting the Krylov wave function spread into an environment of non-local observables.
The only assumption needed to introduce the boundary is that the wave function is continuous at the edge of the Krylov chain. Further details on the construction and interpretation of this boundary condition can be found in Ref. \cite{Loizeau2025}. 

The dynamical symmetries of $H$ are the eigenmodes of $\Tilde{\LL}$ and come in two classes: transient and perpetual depending on the imaginary part of their frequency. 
Indeed, because we have introduced an open boundary condition, $\Tilde{\LL}$ is not hermitian and therefore can admit transient modes that have a complex frequency and decay in time.

{\it The Saw model -- } Now that we have recalled the Krylov space approach to dynamical symmetries, let's introduce a simplified version of model \eqref{eq:H} that can be treated semi-analytically. In this Saw model, instead of introducing a disorder to the magnetic field, we add an alternating $+w,-w$ perturbation:
\begin{align}
    H =& \sum_i \left( X_i X_{i+1} + Y_i Y_{i+1} + Z_i Z_{i+1} \right)\label{eq:Hsaw}\\ 
    &+\sum_i \left(1+w (-1)^i\right)X_i\nonumber
\end{align}

\begin{figure}
\centering
\includegraphics[width=0.48\textwidth]{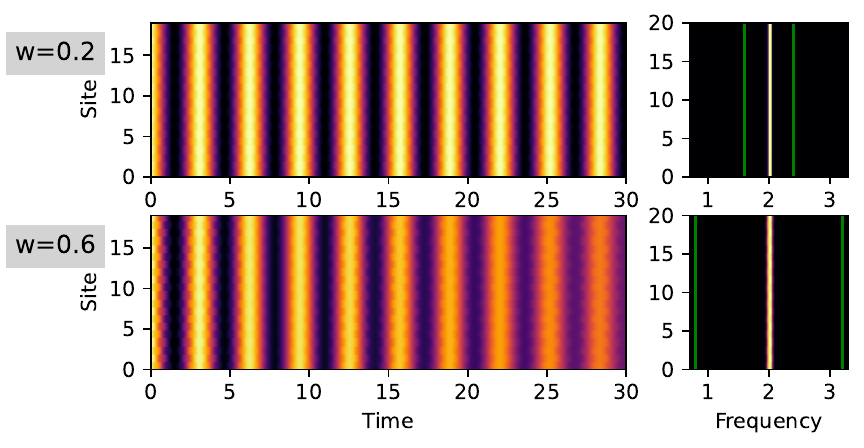}
\caption{Time evolution of $\langle Z_i\rangle$ in the Saw model \eqref{eq:Hsaw}. The green lines on the Fourier plot indicate the two magnetic fields strengths $2(1\pm w)$. We observe strong synchronization: only one frequency $\omega=2$ is visible.}
\label{fig:dynamics_saw}
\end{figure}

The dynamics of this model is extremely robust to disorder, as shown on figure \ref{fig:dynamics_saw}. Even at large disorder (e.g $w=0.6$) the spin dynamics only shows a single frequency.

The Lanczos coefficients of the Saw model can be computed exactly as a function of disorder $W$ for a few orders by running the Lanczos algorithm \eqref{eq:lanczos} symbolically. In practice, we use the Julia package PauliStrings.jl with symbolic coefficients \cite{ps1}.
The first three Lanczos coefficients are
\begin{align}
    b_1 &= 2 \sqrt{w^2+1}= 2+O\left(w^2\right)\\
    b_2 &= \frac{4 \sqrt{3} w}{\sqrt{w^2+1}}=4 \sqrt{3} w+O\left(w^2\right)\\
    b_3 &= \sqrt{\frac{4 \left(3 w^4+22 w^2+55\right)}{3 \left(w^2+1\right)}}\\
    &=2 \sqrt{\frac{55}{3}}+O\left(w^2\right)
\end{align}
(higher orders can be found in the supplements and in ref. \cite{github}).

Remarkably, only $b_2$ has a first order contribution in $w$ and has no zero's order term, meaning it vanishes in the absence of disorder.
Therefore, in the limit $w\to0$, $b_2=0$ and the Krylov chain is exactly cut into two disconnected sectors. When a small $w$ is introduced, then the two sectors are coupled back together. We can treat this as a perturbation of the disorder-free Krylov chain.
Define $b_n'=\lim_{w\to 0} b_n$ the disorder-free Lanczos coefficients, then construct the unperturbed Liouvillian at $w=0$:
\begin{equation}
\LL^0=i\begin{pmatrix}
0   & -b_1'& 0    & 0     & \cdots   \\
b_1' & 0   & 0 & 0     & \cdots   \\
0   & 0 & 0    & -b_3'  & \cdots   \\
0   & 0   & b_3'  & 0     & \cdots   \\
\vdots & \vdots  & \vdots  & \vdots  & 
\end{pmatrix}
\end{equation}
and the perturbation
\begin{equation}
V=4 \sqrt{3} i\begin{pmatrix}
0   & 0   & 0    & 0     & \cdots   \\
0   & 0   & -1 & 0     & \cdots   \\
0   & 1   & 0    & 0  & \cdots   \\
0   & 0   & 0  & 0     & \cdots   \\
\vdots & \vdots  & \vdots  & \vdots  & 
\end{pmatrix}.
\end{equation}
Perturbation $V$ encodes the first order perturbation correction to the Krylov chain in $w$. In only results from the second Lanczos coefficient.

\begin{figure}
\centering
\includegraphics[width=0.48\textwidth]{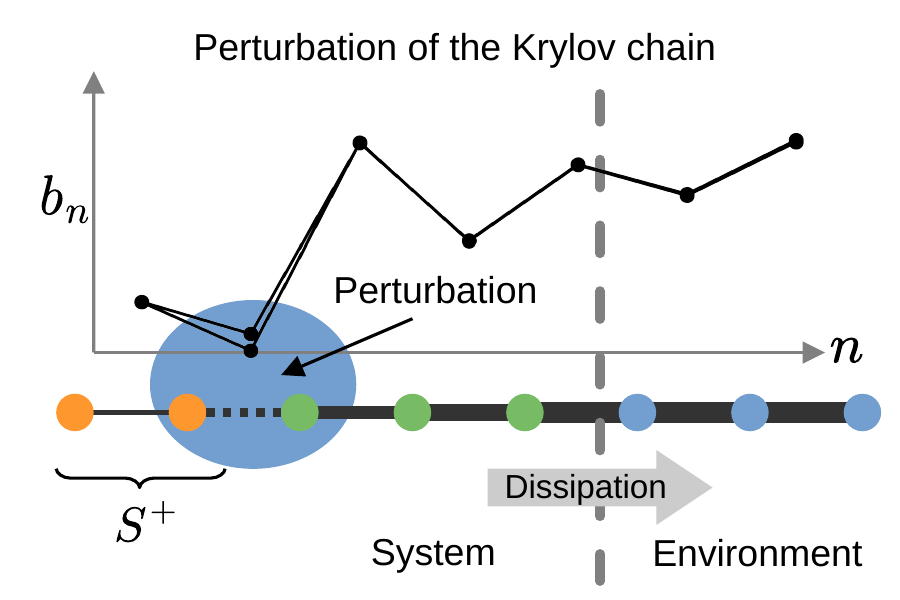}
\caption{In the absence of disorder, the $b_2=0$ and the left of the Krylov chain is inaccessible to the dynamics. When a disorder is introduced, $b_2\sim w^2$ couple the first and second sites of the Krylov chain to the rest of the chain. This can be treated as a perturbation of the Liouvillian. Following ref \cite{Loizeau2025}, we also open the Krylov chain to the right in order to simulate the thermodynamic limit.}
\label{}
\end{figure}

To first order, the perturbed Liouvillian is $\LL'=\LL^0+wV$. We will compute perturbations to $\ket{\psi_2}=\frac{\sqrt 2}{2}(1, i)$, the eigenvector of $\LL^0$ corresponding to $S^+$ in Krylov space.

First note that there is no first order perturbation to $\ket{\psi_2}$. The second order perturbation to the frequency is 
\begin{align}
    E^{(2)}=\sum_{n \neq 2}  \frac{ \bra{\tilde\phi_2}V\ket{\phi_n} \bra{\tilde\phi_n}V\ket{\phi_2}}{E_2-E_n}
\end{align}
\cite{Sternheim1972} where $\ket{\tilde\phi_n}$ and $\ket{\phi_n}$ are the right and left eigenvectors of $\LL^0$. Note that we include the open boundary condition \ref{eq:open_chain} to $\LL^0$, so $\LL^0$ is not hermitian and has complex eigenvalues. The main effect of perturbation $V$ is to couple the left part of the Krylov chain to the rest of the chain. Without the perturbation i.e. when $w=0$, the dynamical symmetry $S^+$ is infinitely long lived because it is confined to the first two sites of the chain. When $V$ is introduced, a probability current can flow through the second Krylov site, effectively introducing a decay rate to the initial dynamical symmetry.

\begin{figure}
\centering
\includegraphics[width=0.48\textwidth]{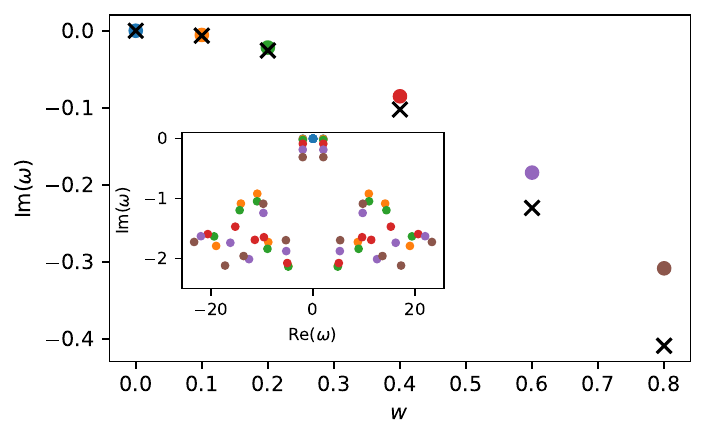}
\caption{Decay rate of the dynamical symmetry v.s disorder $w$ for the Saw model. The colors indicate the exact value and the black crosses indicate the second order perturbation theory result. The inset shows the full spectrum of the Krylov chain for different $w$. Only the perturbed $S^+$ dynamical symmetry is shown on the main plot. This corresponds to the central top dots on the inset. Cf. Ref.~\cite{Loizeau2025} for more details on the method to extract the full Liouvillian spectrum.}
\label{fig:spectrum}
\end{figure}

We can numerically compute $E^{(2)}$ from the $b_n'$s and the modified frequency is $E\approx 2+w^2 E^{(2)}$. This provides direct evidence for synchronization: disorder produces only a second-order correction to the frequency of the dynamical symmetry. By contrast, in the absence of interactions, the frequency would generically acquire a linear dependence on the disorder strength.

In figure \ref{fig:spectrum} we show that for small $w$ we recover the right decay rate of the dynamical symmetry.

{\it Discussion -- }
In this paper we have developed a novel framework for studying the dynamics of perturbed quantum systems based on Liouvillian Krylov space. 
Liouvillian Krylov space provides direct access to the dynamics of a system through its dynamical symmetries.
From this perspective, when analyzing the dynamics generated by a perturbed Hamiltonian $H+V$, it is natural to treat the perturbation within Liouvillian Krylov space rather than in the Hamiltonian Hilbert space. We have shown that, for the synchronized systems considered here, the Krylov-space representation dramatically simplifies the form of the perturbation. The perturbation appears as a coupling between two distinct sectors of the Krylov chain, allowing us to treat the problem analytically.
Using this method, we uncovered a transient dynamical symmetry that decays over time. To our knowledge, this constitutes the first explicit example of an isolated, local transient dynamical symmetry, a dynamical symmetry with a complex valued frequency that decays in time. In Ref.~\cite{Loizeau2025}, we demonstrated that chaotic systems admit a continuum of nonlocal transient dynamical symmetries; however, isolated local instances had not previously been identified.

In future work, we plan to use Lieb-Robinson bounds to characterize the speed of emergence of synchronization, as well as, cases where we only consider a strictly localized seed operator.

Finally, we expect that quantum synchronization and dynamical symmetries will play a central role in understanding subsystem decompositions in quantum theory. 
In quantum theory such subsystems decomposition are not well defined and recent work has focused on developing frameworks that explain how preferred subsystems emerge \cite{Loizeau2023, Loizeau2024mereo, Carroll2021, Soulas2025, Sels2014, Zanardi2024operationalquantum, Carroll2019, Carroll2022, Tegmark2015}.
Local dynamical symmetries can be interpreted as emergent subsystems. When elementary subsystems synchronize, they collectively behave as a new effective subsystem of its own. 
As suggested in Ref.~\cite{Loizeau2024mereo}, the emergence of classicality, and possibly of spacetime itself, might ultimately be a problem about the emergence of preferred tensor product structures or, equivalently, preferred subsystem decompositions. The dynamical symmetries approach provides a promising route toward identifying which degrees of freedom constitute meaningful subsystems in quantum theory.

\vspace{0.5cm}

\begin{acknowledgments}
{\it Data availability -- } A Julia code to symbolically compute the Lanczos coefficients of the Saw model is available at \url{https://github.com/nicolasloizeau/krylov_synchronization} \cite{github}.
\end{acknowledgments}
\vspace{0.5cm}

\begin{acknowledgments}
{\it Acknowledgments -- } We thank L. Mazza, D. Sels and J. Paaske for feedback and useful discussions. 
N.L. and B.B. were supported by a research grant (42085) from Villum Fonden.
B.B. acknowledges funding by the French National Research Agency (ANR) under project ANR-24-CPJ1-0150-01.

We dedicate this work to the memory of M. Medenjak and we are sorry he did not live to see the work completed. B.B. would like to thank his wife Vendi, who is his unsung hero, for all the suggestions and comments on the manuscript. 
\end{acknowledgments}

\bibliography{bib}
\newpage

\section{End matter}
\subsection{Extraction of local dynamical symmetries}
\begin{figure}[H]
\centering
\includegraphics[width=0.49\textwidth]{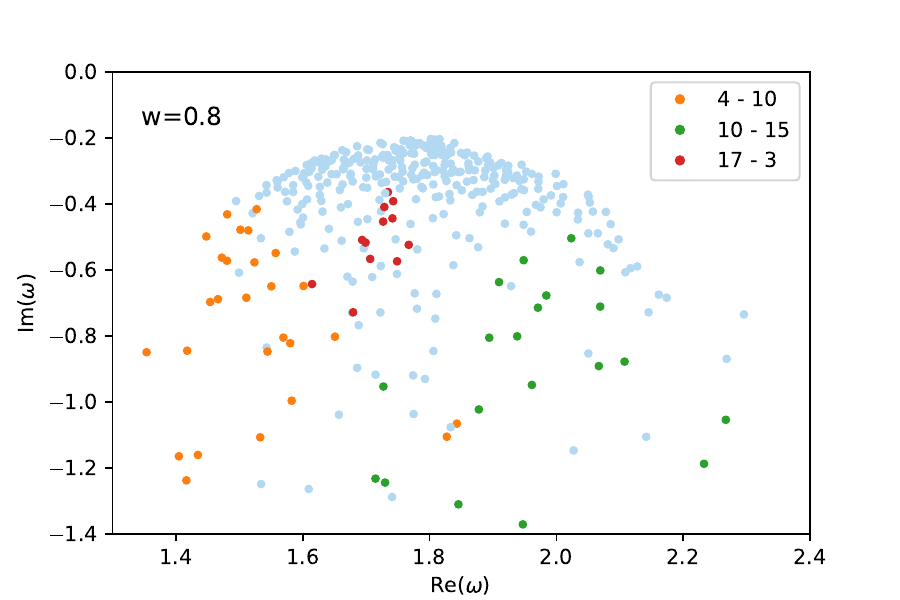}
\caption{Approximate Liouvillian eigenvalues of the disordered Heisenberg model \ref{eq:H} extracted with the Lanczos algorithm. For each data point, we start the Lanczos algorithm with the operator $\sum_{k=i}^j Z_k$. Orange, green and red colors indicate results obtained starting with patches between (i,j)=(4,10), (10,15), (17,3) respectively.}
\label{fig:patches}
\end{figure}

In figure \ref{fig:dynamics} we see that at large disorder (eg w=0.8), spins synchronize with their neighbors, forming local patch that we can attribute to local dynamical symmetries. IN particular, three of these patches appear between spins (4,10), (10,15) and (17,3). In order to check that these correspond to dynamical symmetries, we diagonalize the Liouvillian by constructing the associated Krylov space. However, because translation symmetry is broken, the results are very slow to converge with Krylov dimension and are very sensitive to the initial Lanczos operator $O_0$. To address this, we start the Lanczos algorithm with all $Z$ operators supported on contiguous local patches : $\sum_{k=i}^j Z_k$. In figure \ref{fig:patches} we see that with this strategy we recover the local dynamical symmetries and their frequencies corresponding to the patches seen in figure \ref{fig:dynamics}.

\subsection{Convergence of the Lanczos method}
\begin{figure}[H]
\centering
\includegraphics[width=0.49\textwidth]{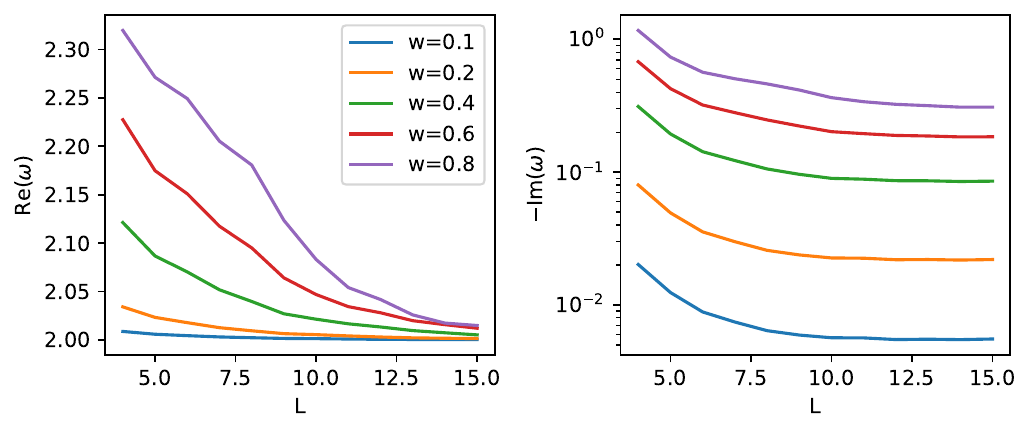}
\caption{Convergence of the real and imaginary part of the dynamical symmetry in the Saw model vs Krylov space dimension (Lanczos steps). The imaginary part converges to a non zero value, leading to a transient dynamical symmetry. These are computed numerically by diagonalizing \ref{eq:open_chain}.}
\label{fig:convergence}
\end{figure}

\vspace{-3cm}
When diagonalizing \eqref{eq:open_chain}, we get complex eigenvaluess meaning that some dynamical symmetries decay over time. In particular in the Saw model, the perturbed $S^+$ dynamical symmetry picks a non-zero imaginary part (cf fig \ref{fig:spectrum}). 
In order to check that this is not a finite size effect, we show on figure \ref{fig:convergence} the convergence of the imaginary part of the perturbed $S^+$ frequency vs Krylov space dimension. We clearly see that the imaginary part of the dynamical symmetry converges to a non-zero value.
\\

\end{document}


\title{Supplemental Material\\Krylov space perturbation theory for quantum synchronization in closed systems}
\author{Nicolas Loizeau}
\affiliation{Niels Bohr Institute, University of Copenhagen, Copenhagen, Denmark}
\author{Berislav Bu\v ca}
\affiliation{Niels Bohr Institute, University of Copenhagen, Copenhagen, Denmark}
\affiliation{Universit\'{e} Paris-Saclay, CNRS, LPTMS, 91405, Orsay, France}
\affiliation{Clarendon Laboratory, University of Oxford, Parks Road, Oxford OX1 3PU, United Kingdom}

\date{\today}
\maketitle

\newpage
\onecolumngrid

\def\theequation{S\arabic{equation}}
\setcounter{equation}{0}


\onecolumngrid

\subsection{Exact Lanczos coefficients for the Saw model}

We compute the exact Lanczos coefficient symbolically using PauliStrings.jl. In practice we can get the first 8 but they are too large to fit here. The code to compute the coefficients is available at \url{https://github.com/nicolasloizeau/krylov_synchronization}.
\begin{align}
b_1^2=&4 \left(W^2+1\right)\\
b_2^2=&\frac{48 W^2}{W^2+1}\\
b_3^2=&\frac{4 \left(3 W^4+22 W^2+55\right)}{3 \left(W^2+1\right)}\\
b_4^2=&\frac{16 \left(84 W^4+293 W^2+209\right)}{9 W^4+66 W^2+165}\\
b_5^2=&\frac{12 \left(468 W^8+4781 W^6+26829 W^4+77275 W^2+55759\right)}{\left(84 W^2+209\right) \left(3 W^4+22 W^2+55\right)}\\
b_6^2=&\frac{16 \left(3 W^4+22 W^2+55\right) \left(9216 W^8+175760 W^6+759424 W^4+1739727 W^2+538097\right)}{\left(84 W^2+209\right) \left(468 W^8+4781 W^6+26829 W^4+77275 W^2+55759\right)}\\
\end{align}
First order expansions in $W$ of the first 8 Lanczos coefficients:
\begin{align}
b_1&=2+O\left(W^2\right)\\
b_2&=4 \sqrt{3} W+O\left(W^2\right)\\
b_3&=2 \sqrt{\frac{55}{3}}+O\left(W^2\right)\\
b_4&=4 \sqrt{\frac{19}{15}}+O\left(W^2\right)\\
b_5&=2 \sqrt{\frac{15207}{1045}}+O\left(W^2\right)\\
b_6&=4\sqrt{\frac{2690485}{1059421}}+O\left(W^2\right)\\
b_7&=\sqrt{\frac{200726656777}{2727613693}}+O\left(W^2\right)\\
b_8&=\sqrt{\frac{346659764178550537}{5684758517459651}}+O\left(
   W^2\right).
\end{align}

\subsection{Additional figures}

\begin{figure}[H]
\centering
\includegraphics[width=0.8\textwidth]{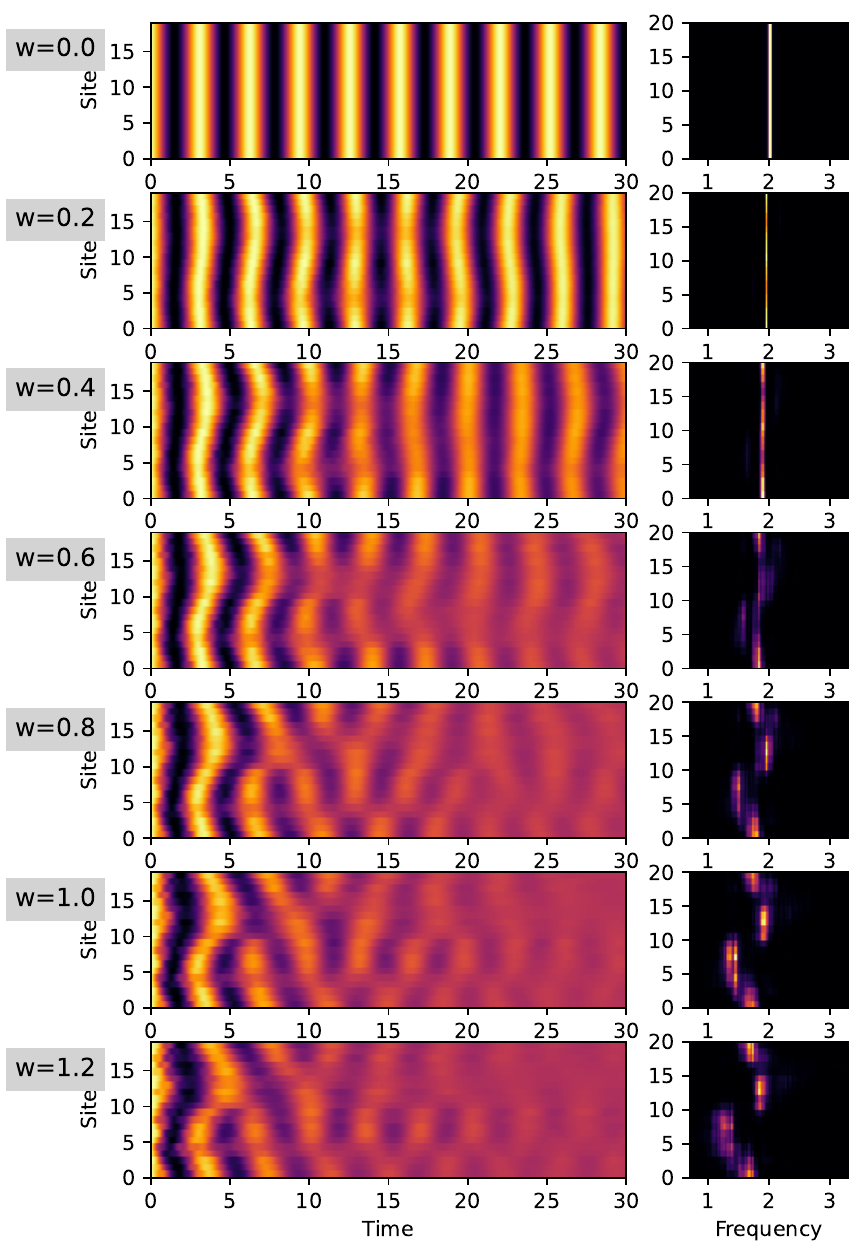}
\caption{Supplement to figure 2. Time evolution of $\langle Z_i\rangle$ for a single random realization of the disordered Heisenberg chain.}
\end{figure}

\begin{figure}[H]
\centering
\includegraphics[width=0.8\textwidth]{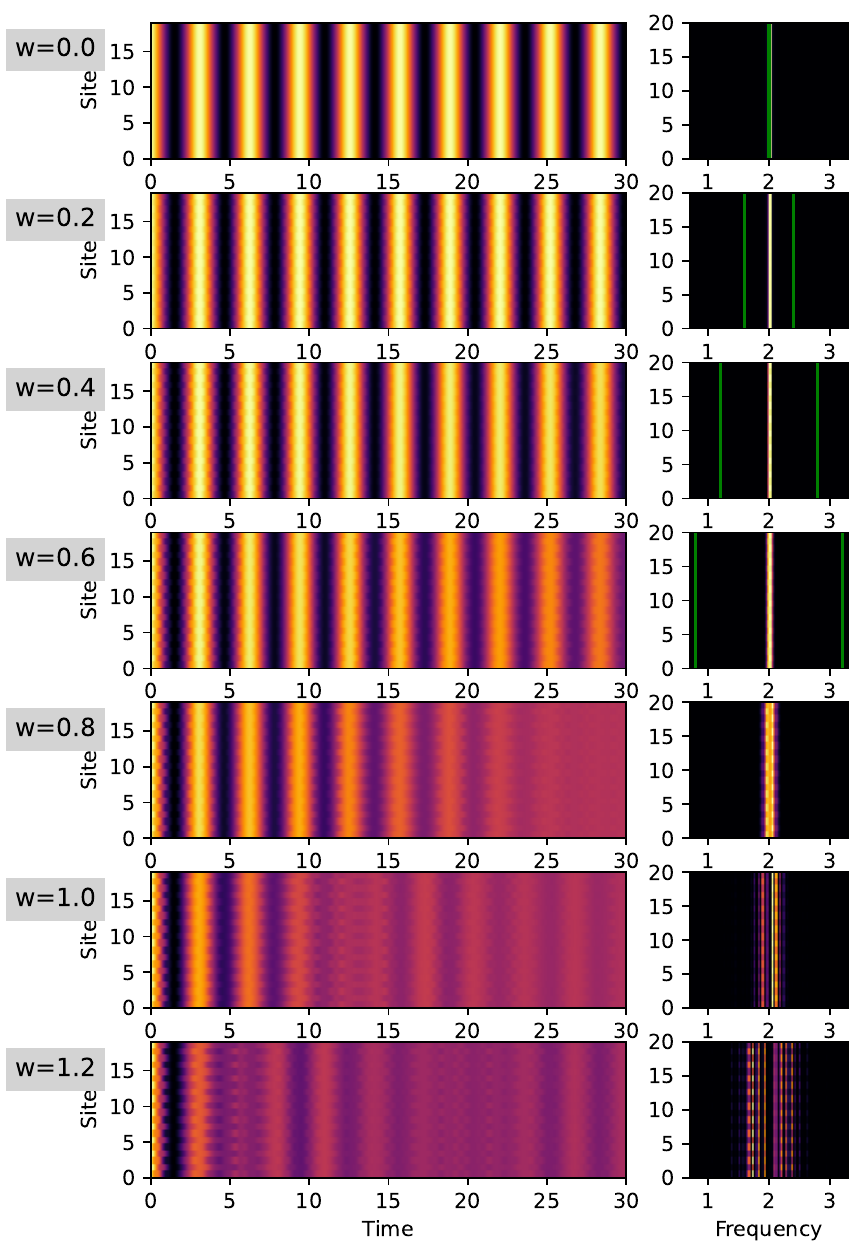}
\caption{Supplement to figure 3. Time evolution of $\langle Z_i\rangle$ in the Saw model.}
\end{figure}